
\input amstex
\magnification=\magstep1
\vsize=23truecm
\hsize=15.5truecm
\hoffset=2.4truecm
\voffset=1.5truecm
\parskip=.2truecm
\font\ut=cmbx10
\font\ti=cmbx10 scaled\magstep1

\font\ninerm=cmr9

\def\de{\delta}

\def\la{\lambda}

\def\pa{\partial}

\def\vphi{\varphi}
\font\ut=cmbx10 scaled\magstep1
\font\ti=cmbx10 scaled\magstep2

\baselineskip=.6truecm
{\hfill ZU-TH-13/92}
\vskip.1truecm
{\hfill March 1992}
\vskip1.0truecm
\centerline{\ti Matter Coupling}
\vskip0.1truecm
\centerline{\ti and Spontaneous Symmetry Breaking}
\vskip0.1truecm
\centerline{\ti in Topological Gravity}
\vskip2.0truecm
\centerline{
T. T. Burwick\,
\plainfootnote{$^{a)}$}
{\ninerm Supported by the Swiss National Science Foundation}
\plainfootnote{$^{b)}$}
{\ninerm Adress after 1 may 1992: SLAC,
Stanford University, Stanford, CA 94309, USA}
and\,A. H. Chamseddine$\, ^{a)}$
}
\vskip0.5truecm
\centerline{\it Institute for Theoretical Physics}
\centerline{\it University of Z\"urich, Sch\"onberggasse 9}
\centerline{\it CH-8001 Z\"urich}
\vskip1.0truecm
\centerline{
K. A. Meissner
}
\vskip0.5truecm
\centerline{\it Institute for Theoretical Physics}
\centerline{\it Warsaw University, ul. Hoza 69}
\centerline{\it 00-681 Warsaw, Poland}

\vskip2.0truecm
\centerline{\bf Abstract}
{\sl
Matter is coupled to three-dimensional gravity
such that the topological phase is allowed and
the (anti-) de Sitter or Poincar\'e symmetry remains intact.
Spontaneous symmetry breaking to the Lorentz group
occurs if a scalar field is included. This Higgs field can then
be used to couple matter so that the familiar form of the matter
coupling is established in the broken phase.
We also give the supersymmetrization of this construction.
}
\vskip.5truecm
\vfill
\eject

\bigskip\centerline{\ut 1. Introduction}
\vskip1truecm
\TagsOnRight
\noindent
Many attempts have been made to formulate a quantum theory of
four-dimen\-sional gravity (see [1]).
In this approach it is hoped that if
gravity can be formulated as a renormalizable theory,
then this would improve
the prospects of unifying gravity with the other known
interactions. The recent
developments in three-dimensional gravity provide an
excellent testing ground
for this program. There it was shown that by formulating
three-dimensional
gravity as a topological gauge theory of the
groups $SO(1,3),\, SO(2,2)$ or
$ISO(1,2)$, the theory becomes finite [2].
The main difficulty in advancing this program is
the coupling of matter.

A first difficulty of introducing matter lies in the
nature of topological gravity. It allows for the unbroken
phase of gravity where the dreibein is
degenerate: $e^a_\mu=0$. In this topological phase, the
notion of geometry loses its meaning. Physics in the
usual sense, where space-time is equipped with distances,
arises only away from this phase. Actually, in [2]
the partition function was seen to be dominated
by geometrical universes, but the appearance of the
topological phase was essential for its derivation.
Matter coupling, however, is usually formulated by using the inverse
dreibein $e^\mu_a$ which would become singular in
the topological phase. In topological gravity this coupling
has to be introduced such that only $e^a_\mu$ is used.
Moreover, we want to require the matter coupling to
reproduce the familiar form if restricted to invertible dreibeins.

A second difficulty stems from the fact that $e^a_\mu$ is part of
the gauge field $A$ and cannot be used by itself without breaking
the gauge invariance. It is then suggestive to break the gauge
symmetry to the Lorentz group $SO(1,2)$ so that $e^a_\mu$
would correspond to the broken symmetry.
To break the symmetry we employ some kind of Higgs mechanism.
However, writing a usual Higgs potential in the action requires
a metric for the volume element. Again, since no dreibein
and therefore no metric $g_{\mu\nu}=e_\mu^a e_{\nu a}$ can be used
without breaking the tangent space symmetry "by hand",
the Higgs field potential terms cannot be written in the usual
way and an alternative construction will be applied.

It is surprising that despite of these difficulties matter
interactions can be introduced. In the following we
will discuss the case of a scalar field.
The plan of this paper is as follows.
In section 2 we give the coupling of
three-dimensional topological gravity to matter.
In section 3 the construction is generalized to the supersymmetric
case.
Some comments and the conclusion are in section 4.

\vskip1.2truecm

\bigskip\centerline{\ut 2. Matter Coupling to three-dimensional
Topological Gravity}
\vskip0.8truecm
\noindent
{}From the work of Witten it is now established that three-dimensional
quantum gravity becomes a finite theory when formulated as a gauge
theory of
$G = SO(1,3),~SO(2,2)$ or $ISO(1,2)$ depending on
the sign of
the cosmological constant [2].
The gauge invariant action is of the Chern-Simons type
$$
S_g = 4k_g \int < AdA+ {2\over 3}A^3>
\tag2.1
$$
where $A$ is an $SO(2,2)$ gauge field
(the other two cases are recovered by Wick rotation or
an In\"onu-Wigner group contraction)
$$
A = {1\over 4} A^{AB}J_{AB}, \quad A = a,3 ;\quad a = 0,1,2
$$
and the quadratic form is defined by
$$
< J_{AB}J_{CD}> = \epsilon_{ABCD}
\tag2.2
$$
The connection with gravity is made through the identification
$$
A^{a3} \equiv e^a,\quad A^{ab} \equiv \omega^{ab}
\tag2.3
$$
In terms of $e$ and the spin connection $\omega$  the
action (2.1) takes the form
$$
S_g = k_g \int \epsilon_{abc}e^a
( R^{bc} -{1\over 3} e^be^c)
\tag2.4
$$
where $R^{bc} = d\omega^{bc} + \omega^{bd}\omega_d^c$.
At the classical level, when $e_{\mu}^a$ is restricted to the
subspace of
invertible fields, the action (2.4) is equivalent to the
Einstein-Hilbert
action. However, this equivalence breaks down at the quantum
level, where the quantum theory of (2.4) is finite.

The main disadvantage in this formulation is the
difficulty of introducing non-trivial matter.
By "non-trivial" we mean couplings which, in the
non-topological phase, reduce to the familiar interactions.
The familiar form of the bosonic matter coupling requires the
inverse dreibein $e^\mu_a$. This, however, is singular in the
topological phase where $e_\mu^a=0$. Moreover, $e_\mu^a$ is part
of the gauge field and cannot be used by itself without breaking
the symmetry.
At the quantum level, the action (2.4) generates divergent 1-loop
diagrams that are cancelled by ghost diagrams arising from
Lorentz- and translation invariance. Since we do not want to
lose these ghost diagrams, we should try to break the gauge
spontaneously.
Let us try to couple a scalar field $H$.
Without using a metric, the only coupling that could be introduced
would be to multiply the action (2.1) by factors of $H$.
This, certainly, does not give interesting physics.
Attempts have been made to introduce an additional
antisymmetric tensor [3] or fields living in representations
of only the Lorentz-group SO(1,2) [4].
These, however, have a trivial physical content.
Here, instead, we take a different strategy. We consider a field $H^A$
and identify $H^3=H$ [5].
We will see that when expanding around a flat background and
using a linear approximation, the $H^A$ coupling will reproduce
the familiar $\pa^\mu H\pa_\mu H$ after eliminating the $H^A$ by
its equation of motion.
Since the $H^A$ will take a non-zero vacuum expectation
value (vev)
which breaks the symmetry to the Lorentz-group,
we call it a Higgs field. This Higgs field can then
be used to couple other matter fields.

Since no metric is at our disposal to write volume elements,
the only Higgs terms that can be written
(apart from possible factors $H^AH_A$ multiplying them) are
$$
S_h = -\int\epsilon_{ABCD} H^A[\mu DH^BF^{CD} + \la DH^BDH^CDH^D]
\tag2.5
$$
where
$$
\alignat 2
D_{\mu}H^A &= &&\,\partial_{\mu}H^A + A_{\mu}^{AB}H_B  \\
F^{AB} &= &&\,dA^{AB} + A^{AC} A_C^B
\endalignat
$$
With the Higgs terms given in (2.5) we may now ask for a possible
vev in the translationary direction:
$$
H^A\,=\,\,<H^A>\,+\,{\bar H}^A
\tag2.6
$$
where
$$
<H^a>\,=\,0\quad,\quad <H^3>\,\equiv\,<H>
\tag2.7
$$
Actually, the argument should have been reversed: It is the
direction of the non-zero $<H^A>$ that decides which part of
the gauge fields in (2.3) seperates to be identified with the dreibein.
To look for non-zero $<H>$ we have to consider the part of the
action (2.5) given by
$$
S_h^\prime = \int\epsilon_{abc}(\mu H^2e^aR^{bc}
+ e^ae^be^c[-\mu H^2+\la H^4])
\tag2.8
$$
The $ ^\prime$ indicates that in (2.5) we set $H^a=0$ which is
sufficient for
obtaining a vev in the translation direction.
With ${1\over 3!}\epsilon_{abc}e^ae^be^c=d^3x\sqrt{g}\,$,
the last two terms in (2.8) are seen to be the usual
scalar potential.
It is a well-known feature (and problem!) that the vev of a Higgs
field changes the cosmological constant.
For convenience, we may assume that we tuned the
coupling constants such that the effective cosmological
constant vanishes. Then we may go to the flat background:
$$
<e_\mu^a>\,=\,\de_\mu^a\quad,\quad <\omega_\mu^{ab}>\,=\,0
\tag2.9
$$
For such a background, the first term in (2.8) will not contribute.
With $\la>0$, the Higgs potential in (2.8) is then minimized by
$$
<H>\,=\,\sqrt{\mu\over 2\la}
\tag2.10
$$
if $\mu>0$, otherwise the vev will vanish.
The vev (2.10) breaks the tangent space
symmetry to the Lorentz-group $SO(1,2)$ that leaves (2.7) invariant.

Plugging (2.10) back into (2.8), we find the total action to be
$$
S_g + S_h = \int \epsilon_{abc}[(k_g+{\mu^2\over 2\la})e^aR^{bc}
-({1\over 3}k_g+{\mu^2\over4\la})e^ae^be^c]
+ O(\bar H^A)
\tag2.11
$$
Except for Higgs quantum fluctuations, this is of the same form
as the gravity action (2.4) but with a different cosmological constant.
We find this effective cosmological constant to be cancelled if
$$
\mu^2\,=\,-{4\over 3}k_g\la
\tag2.12
$$
This allows to use the flat background (2.9), and in the
linear approximation the terms of (2.5) that are
quadratic in $H^A$ are
$$
2\mu\,\int d^3x\,(2H^a\delta_a^\mu\pa_\mu H\,-\,3H^2\,-\,H^aH_a)
\tag2.13
$$
Eliminating the $H^a$ by its equation of motion from (2.13), this
turns into
$$
2\mu\,\int d^3x\,(\,\pa^\mu H\pa_\mu H\,-\,3H^2\,)
\tag2.14
$$
In (2.14) we recognize the usual kinetic term for the Higgs field
around the flat background (2.9).
For a general gravitational background and including also
higher than quadratic terms in (2.13), the elimination of $H^a$
by equations of motion becomes a formidable task and will not
be attacked here. We take (2.14) as sufficient in determining the
structure of the $H^A$ sector.
Alternatively, for analyzing the system (2.5), the gauge condition
$H^a=0$ could be imposed
and a kinetic term for the Higgs field $H$ would be
generated by a Weyl scaling that absorbs the $H^2$
in the first term of (2.8).

Having included the Higgs field $H^A$, we are now able to couple other
matter fields.
The simplest matter interaction to construct is that of a scalar
multiplet. Let $X^A$ be a scalar multiplet in the fundamental
representation
of $SO(2,2)$ with the identifications $X^a = \pi^a, X^3 =\phi$. One
possible action that reproduces the familiar form at the
classical level is
$$
S_m = k_m \int \epsilon_{ABCD} H^A ~DH^B ~DH^C (X^D~DX^E
H_E)
\tag2.15
$$
If we expand the Higgs field around the broken phase (2.7), (2.10)
the matter action (2.15) takes the form
$$
S_m = -k_m^\prime \int d^3x\epsilon^{\mu\nu\rho}\epsilon_{abc}
e^a_{\mu}e^b_{\nu}\pi^c(\partial_{\rho}\phi - e^d_{\rho}\pi_d)
+ O(\bar H^A)
\tag2.16
$$
where $k_m^\prime = k_m\mu^2/4\la^2$.
The action (2.16) is just the first-order formulation of a scalar
field action. To
see this, assume the non-topological phase where $e^a_{\mu}$ is
invertible,
and substitute the  equation of motion of $\pi_a$,
$$
\pi_a = {1\over 2} e_a^{\mu}\partial_{\mu}\phi
+ O(\bar H^A)
\tag2.17
$$
into the action (2.16) to get
$$
S_m = -\, {k_m^\prime\over2}\int d^3x~e~e_a^{\mu}e^{\nu
a}\partial_{\mu}\phi~\partial_{\nu}\phi
+ O(\bar H^A)
\tag2.18
$$
Thus (2.15) reproduces the canonical form at the classical level.
In the spontaneously broken phase, the total action,
which is the sum of (2.11) and (2.16), has only the
$SO(1,2)$ Lorentz symmetry.

\vskip1.2truecm

\bigskip\centerline{\ut 3. Topological Supergravity and Matter Coupling}
\vskip0.8truecm
\noindent
Since $SO(2,2)\cong SO(1,2)\times SO(1,2)$ and $OSP(2\vert 1)$ is the
graded version of
$SO(1,2)$, the supersymmetric analogue of the construction given
in the previous
section is achieved by gauging $OSP(2\mid 1)\times OSP(2\mid 1)$ [2,6].

We shall adopt the notation of [7] for the matrix representation of
$OSP(2\mid 1)$. Let $\Phi_1$ and $\Phi_2$ be the gauge fields of the two
$OSP(2\mid 1)$ gauge groups transforming as
$$
\aligned
 \Phi_1 &\rightarrow\Omega_1\Phi_1\Omega_1^{-1}
 + \Omega_1d\Omega_1^{-1}\cr
 \Phi_2 &\rightarrow\Omega_2\Phi_2\Omega_2^{-1} + \Omega_2d\Omega_2^{-1}
\endaligned
\tag3.1
$$
where $\Omega_1$ and $\Omega_2$ are two elements of the two
respective groups.
These can be represented in the matrix form
$$
\Phi = \left(\matrix
A_{\alpha}^{\beta} & \psi_{\alpha}\\
\bar\psi^{\beta} & 0
\endmatrix\right)
\tag3.2
$$
where
$$
A_{\alpha\beta} = A_{\beta\alpha},\quad \psi_{\alpha} =
\epsilon_{\alpha\beta}
\bar\psi^{\beta}
\tag3.3
$$
It is also convenient to write
$$
A_{\alpha}^{\beta} = A^{a}
(\tau_a)_{\alpha}^{\beta}
$$
where the $\tau_a$ are the $SO(2,1)$-generators
$$
\tau_0 = \frac{1}{2}
\left(\matrix
0 & 1\\
-1 & 0
\endmatrix\right),\quad
\tau_1 = \frac{1}{2}
\left(\matrix
0 & 1\\
1 & 0
\endmatrix\right),\quad
\tau_2 = \frac{1}{2}
\left(\matrix
1 & 0\\
0 & -1
\endmatrix\right)
$$
Introduce now the Higgs field $G$ transforming as
$$
G\rightarrow\Omega_1\, G\, \Omega_2^{-1}
\tag3.4
$$
and the covariant derivative of $G$, transforming as $G$, is defined by
$$
DG = dG + \Phi_1G - G\Phi_2
\tag3.5
$$
In order to distinguish the group indices of the
second $OSP(2\vert 1)$ let us
denote them by $\dot\alpha ,\dot\beta ,\ldots$. Then the
matrix representation of $G$ is
$$
G =
\left(\matrix
H_{\alpha}^{\dot\beta} & \eta_{\alpha}\\
\bar\xi^{\dot\beta} & \phi
\endmatrix\right)
\tag3.6
$$
where both $\eta_{\alpha}$ and $\xi_{\dot\alpha}$ are Majorana
spinors, and $H_{\alpha}^{\dot\beta}$ and $\phi$ are real.

It will also be necessary to define the equivalent
representation $\tilde G$ transforming as
$$
\tilde G\rightarrow\Omega_2\,\tilde G\,\Omega_1^{-1}
\tag3.7
$$
and whose matrix form is
$$
\tilde G = \pmatrix\big[\epsilon
H^T\epsilon^{-1}\big]_{\dot\alpha}^{\beta} & -\xi_{\dot\alpha}\\
-\bar\eta^{\beta} & \phi\endpmatrix
\tag3.8
$$

We first write the pure supergravity action [6]
$$
S_{sg} = -\frac{k_{sg}}{2}\int \big [ Str(\Phi_1d\Phi_1
+ \frac{2}{3}\Phi_1^3) -
1\rightarrow 2 \big ]
\tag3.9
$$
whose component form is
$$
S_{sg} = \frac{k_{sg}}{4}\int \big[(A_{1a}dA_1^a-\frac{1}{3}
\epsilon_{abc}A_1^aA_1^bA_1^c)
+ 4\bar\psi_1D_1\psi_1 - 1\rightarrow 2\big]
\tag3.10
$$
where $D_i = d+A_i$.The action in (3.10) can be put into a more familiar
form by reexpressing it in terms of [6]
$$
\alignat 2
\omega^a &= &&\frac{1}{2}~(A^a_1 + A_2^a)\\
e^a &= &&\frac{1}{2}~(A^a_1 - A^a_2) \tag3.11\\
\psi_{\pm} &= &&\frac{1}{2}~(\psi_1\pm\psi_2)
\endalignat
$$
Then
$$
\align
S_{sg} = k_{sg}
 \int \quad\big[&e^a(R_a
-\frac{1}{6}\epsilon_{abc} e^be^c) \tag3.12\\
&+ 4\bar\psi_-(d+\omega )\psi_+
 + 2\bar\psi_+e\psi_+ + 2\bar\psi_-e\psi_-\Big]
\endalign
$$
where $R_a=d\omega_a -\frac{1}{2}
\epsilon_{abc}\omega^b\omega^c$.
Using $\omega^a =\frac{1}{2}\epsilon^{abc}\omega_{bc}$ the
bosonic part aggrees
with (2.4).

Apart from trace factors $Str(G\tilde G)$,
the most general expression for the Higgs interactions compatible
with (3.9) and the diagonalization in (3.11) is
$$
\align
S_{sh} = \int \quad\{ &\frac{\mu}{2}\big[Str(G{\widetilde{DG}}
(d\Phi_1+\Phi^2_1)) - Str(\tilde G DG(d\Phi_2 + \Phi^2_2))\big]
\\ &+ {\la\over 4} Str(G{\widetilde{DG}}~DG~{\widetilde{DG}})\}
\tag3.13
\endalign
$$
Analogously to the bosonic case, we may look for a non-zero vev of
$$
G\,=\,<G>\,+\,\bar G
\tag3.14
$$
where
$$
<G>\, = \pmatrix <h> & 0 \\ 0 & <\vphi> \endpmatrix
\tag3.15
$$
and $h$ is in the unit direction of $H=h+H^a\tau_a$.
The supergroup $OSP(2\mid 1)\times OSP$
\linebreak
$(2\mid 1)$ has ten
degrees of freedom.
Out of these, seven may be used to rotate $<G>$ into the direction
given by (3.15). Therefore, a non-zero vev (3.15)
would leave only three
degrees of freedom and we will see that these correspond to the
Lorentz-group.
Since we have to vary the action in the direction given by (3.15)
we only need to look at the terms
$$
\align
S_{sh}^\prime =
 \int \quad\{&\mu h^2 e^aR_a
      -\frac{1}{2}\epsilon_{abc} e^ae^be^c(\mu h^2-\la h^4) \\
 &+ 2\mu(h^2+\vphi^2)  \bar\psi_-(d+\omega)\psi_+\tag3.16\\
 &+ [\mu( 3h^2+\vphi^2-2h\vphi )
      -{\la\over 2}(2h^4-5h^3\vphi+4h^2\vphi^2-h\vphi^3)]
                   \bar\psi_+e\psi_+\\
 &+ [\mu( 3h^2+\vphi^2+2h\vphi )
      -{\la\over 2}(2h^4+5h^3\vphi+4h^2\vphi^2+h\vphi^3)]
                   \bar\psi_-e\psi_-\\
 &+ {\la\over 2}(hd\vphi-\vphi dh)(h^2-\vphi^2)\bar\psi_-\psi_+ \}
\endalign
$$
where $ ^\prime$ indicates that we set Higgs components
orthogonal to (3.15) to zero.
Like the bosonic case, we may for convenience assume that
the coefficients $k_{sg}, \mu$ and $\la$ are tuned such that
the effective cosmological constant vanishes. Then we may
go to a flat background:
$$
<e_\mu^a>\,=\,\de_\mu^a\quad,\quad <\omega_\mu^{a}>\,=\,0\quad,\quad
<\psi_\pm>\,=\,0
\tag3.17
$$
With $\la>0$, the potential in (3.16) will then be minimized by
$$
<h>\,=\,\sqrt{\mu\over 2\la}
\tag3.18
$$
if $\mu>0$. For $\mu<0$ the vev of $h$ would be zero.
The field $\vphi$ is not driven to a certain value;
in the background (3.17) any value for the $\vphi$ is allowed.
The total action takes a particularly interesting form
if we shift
$$
\vphi\rightarrow h\,+\,\vphi
\tag3.19
$$
Then the sum of (3.9) and (3.13) becomes
$$
\align
S_{sg} + S_{sh} =
 \int \quad\{&(k_{sg}+\mu h^2)[e^aR_a + 4\bar\psi_-(d+\omega)\psi_+
          + 2\bar\psi_+e\psi_+] \\
 &+ (k_{sg}+3\mu h^2-3\la h^4)[-\frac{1}{6}\epsilon_{abc} e^ae^be^c
                               + 2\bar\psi_-e\psi_-] \\
 &+ O(\vphi,\bar G)\}\tag3.20
\endalign
$$
Except for Higgs quantum fluctuations and with zero $\vphi$,
the terms appearing in the action (3.20) are of the same form
as the original
supergravity action (3.12), but with a different cosmological constant.
We find the cosmological constant to be cancelled if
$$
\mu^2\,=\,-{4\over3}\,k_{sg}\,\la
\tag3.21
$$
This will then allow for the flat background (3.17).
Notice, that in this background the quadratic Higgs terms in (3.13) are
$$
\mu\int d^3x\,[
H^a\delta^\mu_a\pa_\mu h - 3h^2 -{1\over 4}H^aH_a
+ {1\over 2} \bar\eta\tau^\mu\pa_\mu\eta
- {1\over 2} \bar\xi\tau^\mu\pa_\mu\xi
- {3\over 8}(\bar\eta\eta + \bar\xi\xi)
]
\tag3.22
$$
After rescaling $H^a\rightarrow 2H^a$,
the $h$ and $H^a$ terms are of the same form as in (2.13),
and the $\eta, \xi$ terms are of the Dirac type.

The presence of the Higgs field $G$ does now allow to couple another
matter field $X$ which is a multiplet transforming like $G$.
The matrix representation of $X$ is given by
$$
X = \pmatrix
{1\over 2}(\phi + s)\delta_{\alpha}^{\dot\alpha} +
\pi^a(\tau_a)_{\alpha}^{\dot\alpha},
&\lambda_{\alpha} - \chi_{\alpha} \\
\bar\chi^{\dot\alpha} & s\endpmatrix
\tag3.23
$$
A matter interaction which reproduce the bosonic matter interactions
(2.16) is
$$
S_{sm} = k_{sm} \int Str~(\widetilde{DG}\, DG\,
\tilde G\,X) Str (\tilde G D X)
\tag3.24
$$
This can be seen by using the vev (3.15), (3.18).
Then (3.24) reduces to
$$
\align
S_{sm} = &\tag3.25\\
-k_{sm}^\prime&\int(\epsilon_{abc}e^ae^b\pi^c +
4\bar\psi_-\tau_a\psi_-\pi^a - 4\bar\psi_-e\lambda) ~
 (d\phi - e^d\pi_d - 2\bar\psi_-\lambda) + 0(\vphi,\bar G)
\endalign
$$
where we used the shift (3.19) and did not write the
$\vphi$-contributions.
They always appear with gravitinos and will
not influence the bosonic part.
Although this action has the correct bosonic interactions
for $\pi^a$ and
$\phi$, however, $s$ and $\chi$ decouples, and $\lambda$ does
not acquire a propagator.

\vskip1.2truecm

\bigskip\centerline{\ut 4. Conclusions and Comments}
\vskip0.8truecm
\noindent
We have constructed matter interactions coupled to gravity
in a topological way.
The dreibein separates from the other
Poincare or (anti-) de Sitter gauge fields
only by spontaneous symmetry breaking. The
matter coupling was introduced without using the
inverse dreibein, thereby allowing for the
unbroken phase of gravity. This became possible
by including a Higgs field and using a first order formalism.
Restricting to the invertible dreibeins, the matter coupling
takes the familiar form if the equations of motion are used.
We worked out the three-dimensional case, but the
generalization to topological gravity in higher dimensions [8]
is straightforward.
We also presented the supersymmetric analogue of this construction.

Future work should examine the quantum theory of the proposed matter
interaction in a perturbative setting.
Since we included matter only by spontaneous symmetry breaking,
we can immediately deduce that the pure gravity sector remains finite.
Since three-dimensional
gravity is a specific example of a Chern-Simons theory,
the perturbative analysis
may be performed along the lines of [9].
For the case of pure gravity perturbation could be performed
in the unbroken background $e^a_\mu=0$. Having included
matter interactions, the fields used will obtain propagators only if
one expands around some non-zero background. For the case
of pure gravity this expansion and the perturbative analysis
was performed in [10]. With matter many new vertices
and diagrams arise. Apart from questions about non-zero
backgrounds any quantum analysis of a topological theory
requires the introduction of a background metric
to fix the gauge and derive propagators.
For pure gravity, the resulting quantum theory remains
independent of this background metric [2,11].
In general, however, it is not guaranteed that a theory
which is metric independent at the classical level
would remain so at the quantum level [12].
It is then of interest to study whether the
property of metric independence is lost in the presence of matter.
We want to emphasize that for proving a possible
metric dependence it is not enough to find divergences
that can only be cancelled by using the background metric.
The situation may be compared to Yang-Mills theory
in the axial gauge $n^\mu A_\mu=0$ where $n^\mu$
plays the role of the background metric.
There, at the one-loop level counterterms have been
found that were dependent on $n^\mu$ [13].
Later, the situation was re-investigated by using BRST methods
and it became possible to control this gauge dependence [14].
A BRST analysis along the lines of [14]
should also be applied to study a possible background
dependence of topological gravity in the presence of matter.
This will then decide whether topological gravity
keeps all of its nice features after matter is coupled.

\vskip1.2truecm

\bigskip
\centerline{\ut References}
\vskip0.8truecm
\item{[1]}  P. van Nieuwenhuizen, {\it Phys. Rep.} {\bf 68} (1981) 191,
            and references therein.

\item{[2]}  E. Witten, {\it Nucl.Phys.} {\bf B311}
            (1988) 96;  {\bf B323} (1989) 113.

\item{[3]}  J. Gegenberg, G. Kunstatter and H.P. Leivo,
            {\it Phys.Lett.} {\bf B252} (1990) 381.

\item{[4]}  S. Carlip and J. Gegenberg, Phys. Rev. D44 (1991) 424.

\item{[5]}  A.H. Chamseddine, {\it Anal.Phys.} {\bf 113} (1978) 219; {\it
            Nucl.Phys.} {\bf B131} (1977) 494; \newline
            H.G. Pagels, {\it Phys.Rev.} {\bf D27} (1983) 2299.

\item{[6]}  A. Ach\'ucarro and P.K. Townsend, {\it Phys.Lett.} {\bf B180}
            (1986) 89.

\item{[7]}  A.H. Chamseddine, A. Salam and T. Strathdee, {\it Nucl.Phys.}
            {\bf B136} (1978) 248.

\item{[8]}  A.H. Chamseddine, {\it Nucl.Phys.} {\bf B346}
            (1990) 213.

\item{[9]} L. Alvarez Gaum\'e, J. Labastida and A.V. Ramallo,
           {\it Nucl.Phys.} {\bf B334} \newline (1990) 103 ; \newline
           E. Guadagnini, M. Martellini and M. Mintchev, {\it
           Phys.Lett.} {\bf B334} (1989) 111; \newline
           C.P. Martin, {\it Phys.Lett.} {\bf B241} (1990) 513;

\item{[10]} S. Deser, J. McCarthy and Z. Yang, {\it Phys.Lett.} {\bf
           B222} (1989) 61.

\item{[11]} D. Ray and I. Singer, {\it Adv.Math.} {\bf 7}
           (1971) 145; {\it Ann.Math.} {\bf 98} (1973) 154.

\item{[12]} M. Blau and G. Thompson, {\it Phys.Lett.}
            {\bf B255} (1991) 535.

\item{[13]} D. M. Capper and G. Leibbrandt,
            {\sl Phys. Rev. \bf D 25} (1982) 1002;
            {\sl Phys. Rev. \bf D 25} (1982) 1009;

\item{[14]} P. Gaigg, O. Piguet, A. Rebhahn and M. Schweda,
            {\sl Phys. Lett.\bf 175 B} (1986) 53.

\end